\batchmode
\documentclass[twocolumn,english]{IEEEtran}
\usepackage[T1]{fontenc}
\usepackage{color}
\usepackage{babel}
\usepackage{amsmath}
\usepackage{amssymb}
\usepackage{graphicx}
\usepackage[unicode=true,
 bookmarks=true,bookmarksnumbered=true,bookmarksopen=true,bookmarksopenlevel=1,
 breaklinks=false,pdfborder={0 0 0},backref=false,colorlinks=false]
 {hyperref}
\hypersetup{pdftitle={Your Title},
 pdfauthor={Your Name},
 pdfpagelayout=OneColumn, pdfnewwindow=true, pdfstartview=XYZ, plainpages=false}

\makeatletter
 \let\oldforeign@language\foreign@language
 \DeclareRobustCommand{\foreign@language}[1]{%
   \lowercase{\oldforeign@language{#1}}}

\ifCLASSOPTIONcompsoc
\usepackage[caption=false,font=normalsize,labelfont=sf,textfont=sf]{subfig}
\else
\usepackage[caption=false,font=footnotesize]{subfig}
\fi
\usepackage[numbers,sort&compress]{natbib}

\makeatother

\begin{document}
\pagestyle{empty}

\title{\textcolor{black}{Analysis of Interference Correlation in Non-Poisson
Networks}}

\author{\IEEEauthorblockN{Juan~Wen$^{\dagger}$$^{\ddagger}$,~Min Sheng$^{\dagger}$,~Kaibin
Huang$^{\ddagger}$,~ Jiandong Li$^{\dagger}$\\
}\IEEEauthorblockA{$^{\dagger}$State Key Laboratory of Integrated
Service Networks, Xidian University, Xi'an, Shaanxi, China\\
$^{\ddagger}$Department of Electrical and Electronic Engineering,
The University of Hong Kong, Hong Kong\\
Email: juanwen66@gmail.com, \{msheng, jdli\}@mail.xidian.edu.cn, huangkb@eee.hku.hk}%
\thanks{This work has been supported by National Natural Science Foundation
of China (61231008, 61172079, 61201141, 61301176, 91338114), 111 Project
(B08038), and 863 project (No.2014AA01A701). %
}}

\markboth{}{Your Name \MakeLowercase{\emph{et al.}}: Your Title}
\maketitle\thispagestyle{empty}
\begin{abstract}
\textcolor{black}{The correlation of interference has been well quantified in Poisson networks where the interferers are independent of each other. However, there exists dependence among the base stations (BSs) in wireless networks. In view of this, we quantify the interference correlation in non-Poisson networks where the interferers are distributed as a Matern cluster process (MCP) and a second-order cluster process (SOCP). Interestingly, it is found  that the correlation coefficient of interference for the Matern
cluster networks,   $\zeta_{MCP}$,  is equal to that for  second-order cluster networks,  $\zeta_{SOCP}$. Furthermore,  they are greater than their counterpart for the  Poisson networks. This  shows that clustering in interferers enhances  the interference correlation. In addition, we show that the correlation coefficients $\zeta_{MCP}$ and $\zeta_{SOCP}$ increase as the average number of points in each cluster,  $c$,  grows, but  decrease with the increase in the cluster radius,  $R$.
More importantly, we point that the effects of clustering on interference correlation can be neglected  as  $\frac{c}{\pi^{2}R^{2}}\rightarrow0$. Finally,  the analytical results are validated by simulations. }\end{abstract}

\section{Introduction}

\IEEEPARstart{I}{nterferers}\textcolor{black}{{}  in wireless networks are both temporally and spatially correlated since they are
subject to finite mobility and real network deployment. This leads to the correlation in  interference \cite{InterfCorreThreeSources,InterfereCorrelationCharacteringI,InterferencePrediction}.
Such correlation significantly affects the performance of systems with retransmission schemes, cooperative relaying, and multiple antennas \cite{MRC}. Thus, it is important to quantify the interference correlation. However, traditional analysis focuses on the interference correlation caused by the temporally correlated interferers.}

Ganti and Haenggi are among the first to  quantify  the interference   correlation in Poisson networks in terms of correlation coefficient \textcolor{black}{\cite{InterferenceCorreLetter}}. They found that there exists spatio-temporal interference correlation in wireless networks due to the slow node mobility. Such
correlation reduces the diversity of networks with retransmissions \cite{IntefCorrDiversityPolynomials} or multi-antenna receivers \cite{InterfCorreDiversityLoss},
and thus degrades the corresponding network performance. Increasing node mobility \cite{CorrelationMobileRandomNet}, the randomness in
fading \textcolor{black}{\cite{InterferenceCorreLetter}} and MAC protocols \cite{LocalDelay_MAC} can reduce the interference correlation. Note that prior analysis  was  conducted on one-tier Poisson networks. The authors in \cite{InterfCorre15globecom,InterfCorre_HCNs} investigated the interference correlation in heterogeneous cellular networks (HCNs) where the BSs follow multiple independent Poisson point processes (PPPs). Except for the aforementioned research, prior work   related to  Poisson networks \cite{InterferenceCorreLetter,IntefCorrDiversityPolynomials,InterfCorreDiversityLoss,CorrelationMobileRandomNet,LocalDelay_MAC,InterfCorre15globecom}
only addresses  the interference correlation caused by the temporally correlated interferers, since the  assumption of Poisson distribution inherently neglects the spatial correlation in BS locations.

There exists dependence including clustering and repulsion among the BS locations, especially for HCNs. Specifically, low power nodes are
always allocated in groups to provide high capacity in hotspots. Furthermore,  they are deployed in the annular region of macrocells to avoid severe
inter-tier interference \cite{HCN_HolosticView,SmallCellDeployment}. We call these two phenomenons as intra-tier dependence (i.e., the
clustering among the BSs within the same tier) and inter-tier dependence (i.e., the repulsion among the BSs belonging to different tiers),
respectively. Considering  the intra-tier dependence, a Matern cluster process (MCP) is promising for modeling the clustered low-power BSs in HCNs due to its analytical tractability \cite{PoissonClusterProcess_TIT,HCN_PoissonCluster,HCN-Dependence}. Given  both the intra- and inter-tier dependence, the authors in
\cite{HCN_SOCP} modeled the low-power BSs as a second-order cluster process (SOCP) and analyzed the corresponding network performance. The resultant spatial correlation among the BSs makes the interference correlation more complex, which has not been investigated in the literature.

In this paper, we quantify the interference correlation caused by the spatial correlation of interferers in terms of interference correlation
coefficient. For simplicity, let $\zeta_{PPP}$, $\zeta_{MCP}$, and $\zeta_{SOCP}$ denote the spatio-temporal interference correlation
coefficients in the cases where the interferers are distributed as a  PPP, MCP, and
SOCP, respectively. The mean numbers of points in each cluster of MCP and SOCP are represented as $c$ and $c_{2}$, respectively. Moreover,
the radiuses  of a typical  cluster of  MCP and SOCP are denoted as $R$ and
$D_{2}$, respectively. The main contributions of this paper are summarized
as follows.
\begin{itemize}
\item The interference correlation coefficients,  $\zeta_{MCP}$ and $\zeta_{SOCP}$, are derived and shown to be greater than $\zeta_{PPP}$. This implies
that there exist positive contributions of clustering on interference correlation. In particular, the contributions of clustering in these
two cases are the same, i.e.,  $\zeta_{MCP}=\zeta_{SOCP}$, if $c=c_{2}$
and $R=D_{2}$.
\item The  correlation coefficients $\zeta_{MCP}$ and $\zeta_{SOCP}$ are found to be  significantly affected by the average number of points in each
cluster,  $c$,  and the radius of a typical  cluster,  $R$. Moreover, it is proved that increasing $c$ or decreasing $R$ enhances the interference
correlation, and vice versa. In addition,  $\zeta_{MCP}$ and $\zeta_{SOCP}$ are  approximately equal  to $\zeta_{PPP}$ in the case of \textcolor{black}{$\frac{c}{\pi^{2}R^{2}}\rightarrow0$.}
\end{itemize}

\begin{figure*}[tp]
\begin{minipage}[t]{0.45\textwidth}%
\includegraphics[scale=0.4]{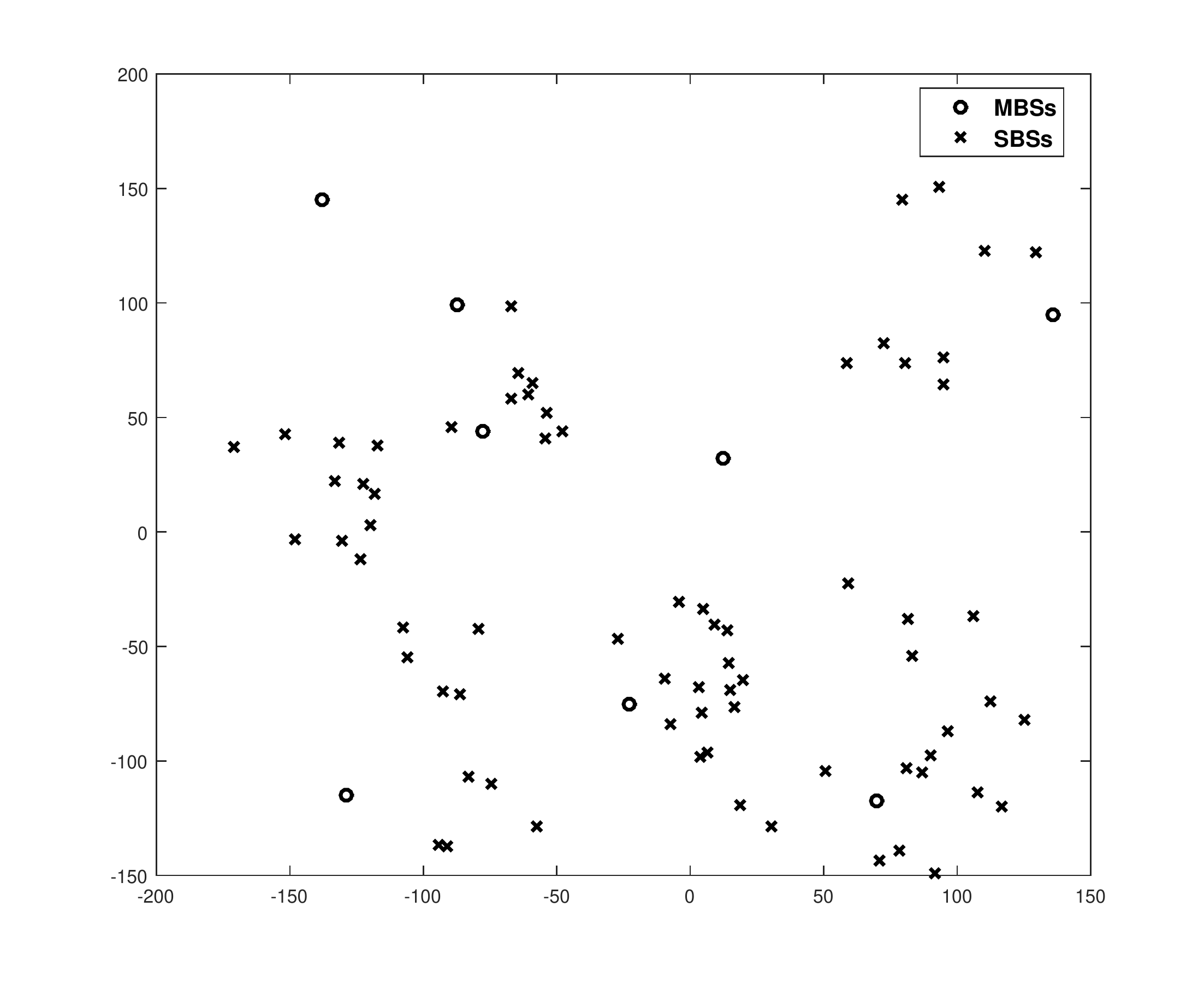}\centering\caption{The two-tier HCN model with intra-tier dependence ($\lambda_{m}=$0.0001,
$\lambda_{p}=0.0002$, $c$=4, $R=30$)}
\end{minipage}\hspace{30bp} %
\begin{minipage}[t]{0.45\textwidth}%
\includegraphics[scale=0.4]{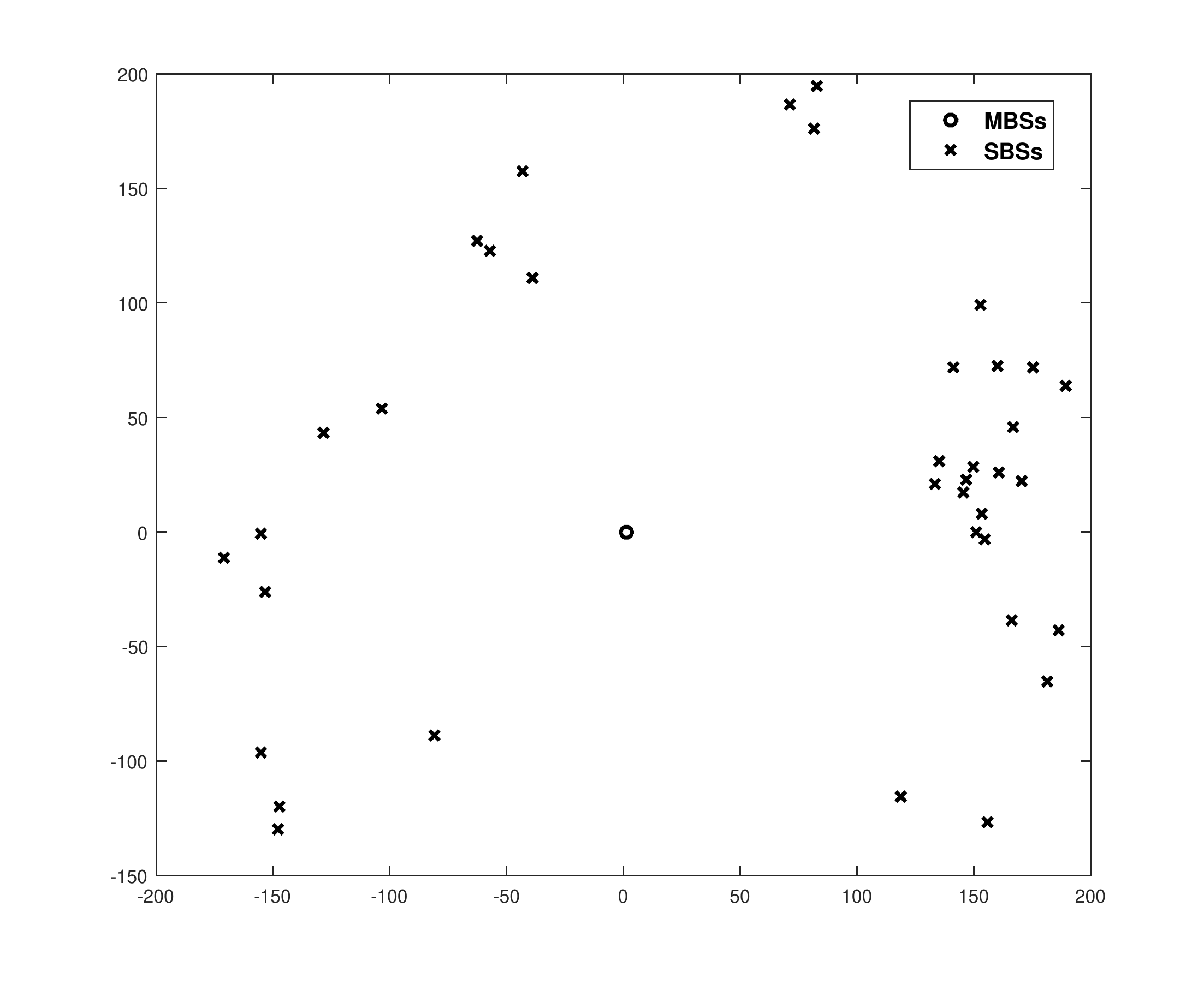}\centering\caption{The two-tier HCN model with both inter-tier and intra-tier dependence
($\lambda_{0}=0.00001,$ $c_{1}=10$, $D_{1}=200$, $c_{2}=3$, $D_{2}=25$)}
\end{minipage}
\end{figure*}

\section{System Model }

We consider a two-tier HCN consisting of macrocell BSs (MBSs) and
small cell BSs (SBSs). Regarding the dependence among the BSs in HCNs,
MBSs and SBSs are respectively modeled as a homogenous PPP and a non-Poisson point process, which is shown in the
sequel. MBSs and SBSs transmit at fixed power $P{}_{tm}$ and $P{}_{ts}$,
respectively. The power received by the user located at $u$ from BS $x$
in time slot $t$ is expressed as $P_{t}h_{xu}\left(t\right)g\left(x-u\right)$,
where $P_{t}$ is $P_{tm}$ or $P_{ts}$ based on the type of BS, $h_{xu}\left(t\right)$
denotes the temporally and spatially independent small-scale fading,
and $g\left(x\right)$ represents the large-scale path loss.

\subsection{Two-tier HCNs Model with Intra-tier Dependence (Case 1)}

In view of intra-tier dependence, SBSs are independently clustered
in Case 1. As shown in Fig. 1, MBSs and SBSs are distributed as a PPP $\phi_{m}=\{x_{1},x_{2},\cdots\}\subset\mathbb{R}^{2}$
with density $\lambda_{m}$ and an independent MCP $\phi_{s}=\left\{ y_{1},y_{2},\cdots\right\} \subset\mathbb{R}^{2}$
whose parent point process is an independent PPP with density $\lambda_{p}$.
To be specific, the number of SBSs in each cluster is a Poisson random variable
with mean $c$. In addition, each SBS is uniformly scattered in a ball
of radius $R$ around its parent point. Thus, the probability density
function (pdf) of the typical cluster whose parent point located at origin is expressed
as
\begin{equation}
f_{cl}\left(y\right)=\begin{cases}
\frac{1}{\pi R^{2}}, & \parallel y\parallel\leq R\\
0, & otherwise.
\end{cases}
\end{equation}

\subsection{Two-tier HCNs Model with both Inter-tier and Intra-tier Dependence
(Case 2)}

As illustrated in Fig. 2%
\footnote{The scenario of multiple macrocells is considered in this paper although only one macrocell is shown for simplicity.%
}, SBSs in Case 2 are clustered in the edge region of macrocell due
to the inter-tier dependence. In Case 2, MBSs follow a PPP $\phi_{m}=\{x_{1},x_{2},\cdots\}\subset\mathbb{R}^{2}$
with density $\lambda_{m}$ and SBSs follow a SOCP $\phi_{s}=\left\{ z_{1},z_{2},\cdots\right\} \subset\mathbb{R}^{2}$
whose parent process is a PPP with density $\lambda_{0}$. These parent
points denote the centers of first-order clusters in which the number
of points is Poisson random variable with mean $c_{1}$. It is worth noting
that in Case 2, MBSs are the parent points of SBSs following SOCP
because of the inter-tier dependence. Thus, $\lambda_{0}=\lambda_{m}$.
Moreover, each point of a first-order cluster is isotropically scattered
according to a centered reverse Gaussian distribution. Thus, the pdf of first-order cluster point
is correspondingly given as \cite{HCN_SOCP}:
\begin{equation}
f_{cl_{1}}\left(x\right)=\begin{cases}
\frac{\left(1-\exp\left(\frac{-\Vert x\Vert^{2}}{2\sigma^{2}}\right)\right)}{\pi D_{1}^{2}+2\pi\sigma^{2}\left(\exp\left(\frac{-D_{1}^{2}}{2\sigma^{2}}\right)-1\right)}, & \Vert x\Vert\leq D_{1}\\
0, & otherwise,
\end{cases}
\end{equation}
where $D_{1}$ is the radius of the coverage area of first-order cluster
and $\sigma$ denotes the standard deviation of reverse Gaussian distribution.
These first-order cluster points are the centers of the second-order
cluster points, i.e., SBSs, whose number is a Poisson random variable with
mean $c_{2}$. Each SBS is uniformly scattered in a ball of radius $D_{2}$
around the first-order cluster points and the corresponding pdf is
given by
\begin{equation}
f_{cl_{2}}\left(x\right)=\begin{cases}
\frac{1}{\pi D_{2}^{2}}, & \parallel x\parallel\leq D_{2}\\
0, & otherwise.
\end{cases}
\end{equation}

\section{Spatio-temporal Interference Correlation}

Since the interference correlation in Poisson networks has been investigated
in \cite{InterferenceCorreLetter}, in this section, we focus on the
spatio-temporal interference correlation when
the interferers follow MCP and SOCP.  We find that the interference
correlations in Matern cluster networks and second-order cluster networks
are greater than that in Poisson networks. Theorem 1 shows the expressions
of these interference correlation coefficients.

\textbf{Theorem 1.} \emph{The spatio-temporal correlation coefficients
of interference when the interferers follow MCP and SOCP are respectively:}
\begin{equation}
\zeta_{MCP}=\frac{\int_{\mathbb{R}^{2}}g\left(x\right)g\left(x-\parallel u-v\parallel\right)\mathrm{d}x+F\left(c,R\right)}{\frac{\mathbb{E}\left[h^{2}\right]}{\mathbb{E}\left[h\right]^{2}}\int_{\mathbb{R}^{2}}g^{2}\left(x\right)\mathrm{d}x+F\left(c,R\right)},\label{eq:CorrCoef_MCP}
\end{equation}
\emph{and}
\begin{equation}
\zeta_{SOCP}=\frac{\int_{\mathbb{R}^{2}}g\left(x\right)g\left(x-\parallel u-v\parallel\right)\mathrm{d}x+F\left(c_{2},D_{2}\right)}{\frac{\mathbb{E}\left[h^{2}\right]}{\mathbb{E}\left[h\right]^{2}}\int_{\mathbb{R}^{2}}g^{2}\left(x\right)\mathrm{d}x+F\left(c_{2},D_{2}\right)},\label{eq:CorrCoef_SOCP}
\end{equation}
\emph{where} $F\left(c,R\right)=\frac{c}{\pi^{2}R^{4}}\int_{\mathbb{R}^{2}}\int_{\mathbb{R}^{2}}g\left(x\right)g\left(y\right)A_{R}\left(\Vert x-y\Vert\right)\mathrm{d}x\mathrm{d}y$,
\emph{and} $A_{R}\left(r\right)=2R^{2}\mathcal{\arccos}\left(\frac{r}{2R}\right)-r\sqrt{R^{2}-\frac{r^{2}}{4}},\;0\leq r\leq2R$,
\emph{and} 0 \emph{for} $r>2R$.
\begin{IEEEproof}
See Appendix A.
\end{IEEEproof}
To better understand the proof, we now give a brief introduction of
the main steps. We calculate the correlation coefficient of interference
according to its definition. Specifically, we first calculate the
mean of interference $I_{t}\left(u\right)$, the mean product of $I_{t_{1}}\left(u\right)$
and $I_{t_{2}}\left(v\right)$, and the second moment of the interference
$I_{t}\left(u\right)$. Then, we substitute the corresponding results
into the definition of interference correlation coefficient to get
the expression.

Recall that the interference correlation coefficient in Poisson networks
is given in \cite{InterferenceCorreLetter} as
\begin{equation}
\zeta_{PPP}=\frac{\int_{\mathbb{R}^{2}}g\left(x\right)g\left(x-\parallel u-v\parallel\right)\mathrm{d}x}{\frac{\mathbb{E}\left[h^{2}\right]}{\mathbb{E}\left[h\right]^{2}}\int_{\mathbb{R}^{2}}g^{2}\left(x\right)\mathrm{d}x}.\label{eq:CorrCoef_PPP}
\end{equation}
Comparing $\zeta_{MCP}$ and $\zeta_{SOCP}$ with $\zeta_{PPP}$,
the function $F\left(c,R\right)$ captures the contribution of clustering
in interferers on interference correlation, which depends on the mean
number of points and the radius of each cluster.

\textbf{Remark 1.} The interference correlation coefficient under
MCP distributed interferers is the same with that under SOCP distributed
interferers if they have the same average number of points and the
radius for each cluster, i.e., $\zeta_{MCP}=\zeta_{SOCP}$ if $c=c_{2}$ and $ R=D_{2}$.
The fact is intuitive because the contribution of the clustering in
interference correlation $F\left(c,R\right)$ is only decided by $c$
and $R$. Further, we give the relationship between the interference
correlation coefficients under MCP or SOCP distributed interferers and
that under PPP distributed interferers in Proposition 1.

\textbf{Proposition 1. }The interference correlation coefficients
$\zeta_{MCP}$ and $\zeta_{SOCP}$ are greater than that in Poisson
networks $\zeta_{PPP}$, i.e., $\zeta_{MCP}>\zeta_{PPP}$ and $\zeta_{SOCP}>\zeta_{PPP}$.
\begin{IEEEproof}
For simplicity, we denote $P=\int_{\mathbb{R}^{2}}g\left(x\right)g\left(x-\parallel u-v\parallel\right)\mathrm{d}x$,
$Q=\frac{\mathbb{E}\left[h^{2}\right]}{\mathbb{E}\left[h\right]^{2}}\int_{\mathbb{R}^{2}}g^{2}\left(x\right)\mathrm{d}x$.
In this way, $\zeta_{PPP}$, $\zeta_{MCP}$, and $\zeta_{SOCP}$ are
simplified to $\zeta_{PPP}=\frac{P}{Q}$, where $P>0$ and $Q>0$, $\zeta_{MCP}=\frac{P+F\left(c,R\right)}{Q+F\left(c,R\right)}$,
and $\zeta_{SOCP}=\frac{P+F\left(c_{2},D_{2}\right)}{Q+F\left(c_{2},D_{2}\right)}$.
Now, we show that $\zeta_{MCP}>\zeta_{PPP}$.
\begin{flalign}
 & \zeta_{MCP}-\zeta_{PPP}=\frac{P+F\left(c,R\right)}{Q+F\left(c,R\right)}-\frac{P}{Q}\nonumber \\
 & =\frac{\left(Q-P\right)F\left(c,R\right)}{Q\left(Q+F\left(c,R\right)\right)}\overset{\left(a\right)}{>0},
\end{flalign}
where $(a)$ comes from the fact that $F\left(c,R\right)>0$ and $Q-P>0$
(since $0<\zeta_{PPP}<1$).

Following the similar steps, we can get $\zeta_{SOCP}>\zeta_{PPP}$.
\end{IEEEproof}
From Proposition 1, we find that the spatial correlation in interferers
significantly affects the correlation in interference. In particular,
the attraction between the interferers enhances the correlation
in interference. Thus, we infer that reducing the clustering in interferers
enables weaken the interference correlation. Next, we study the effect
of system parameters, such as $c$ and $R$, on the interference correlation.

\textbf{Proposition 2.} The interference correlation coefficients
$\zeta_{MCP}$ and $\zeta_{SOCP}$ increase with the increase in $c$
or $c_{2}$, while decrease with the increase in $R$ or $D_{2}$.
In particular, $\zeta_{MCP}\rightarrow\zeta_{PPP}$, if $\frac{c}{\pi^{2}R^{2}}\rightarrow0$
and $\zeta_{SOCP}\rightarrow\zeta_{PPP}$, if $\frac{c_{2}}{\pi^{2}D_{2}^{2}}\rightarrow0$.
\begin{IEEEproof}
Recall that the interference correlation coefficient $\zeta_{MCP}$
can be denoted as $\zeta_{MCP}=\frac{P+F}{Q+F}$. Taking the derivative
of $\zeta_{MCP}$ with respect to $F$, we get $\zeta'_{MCP}=\frac{Q-P}{\left(Q+P\right)^{2}}$.
It should be noted that $0<\zeta_{PPP}=\frac{P}{Q}<1$. Thus, $\zeta'_{MCP}>0$,
i.e., $\zeta_{MCP}$ increases with the increase in $F$. Moreover,
from the expression of $F\left(c,R\right)$, we know that $F\left(c,R\right)\propto c$
and $F\left(c,R\right)\propto\frac{1}{R^{2}}$. As a result, $\zeta_{MCP}$
increases as $c$ increases, while it decreases as $R$ increases.

In addition, $F\left(c,R\right)\rightarrow0$, if $\frac{c}{\pi^{2}R^{2}}\rightarrow0$.
Thus, $\zeta_{MCP}\rightarrow\zeta_{PPP}$, if $\frac{c}{\pi^{2}R^{2}}\rightarrow0$.

The same conclusion for $\zeta_{SOCP}\left(I_{t_{1}}\left(u\right),I_{t_{2}}\left(v\right)\right)$
can be obtained via following the similar steps.
\end{IEEEproof}
The conclusion in Proposition 2 can be explained as follows. Given
$R$ or $D_{2}$, the larger $c$ or $c_{2}$ is, the more attraction
the interferers has. Moreover, the attraction of the interferers has
a positive effect on interference correlation. As a result, the correlation
coefficient increases with the increase in $c$ or $c_{2}$. Similarly,
the correlation coefficient increases with the decrease in $R$ or
$D_{2}$, since the attraction among the interferers becomes strong
when $R$ or $D_{2}$ decreases.

When $\frac{c}{\pi^{2}R^{2}}\rightarrow0$, the effect of clustering
in interferers can be ignored. In particular, when the radius of cluster
$R$ or $D_{2}$ approximates to infinity, MCP and SOCP can be viewed
as the superposition of finite independent PPP. Thus, $\zeta_{MCP}\rightarrow\zeta_{PPP},\: R\rightarrow\infty$
and $\zeta_{SOCP}\rightarrow\zeta_{PPP},\: D_{2}\rightarrow\infty$,
since the summation of multiple independent PPP is also a PPP.

\begin{figure}[t]
\includegraphics[scale=0.4]{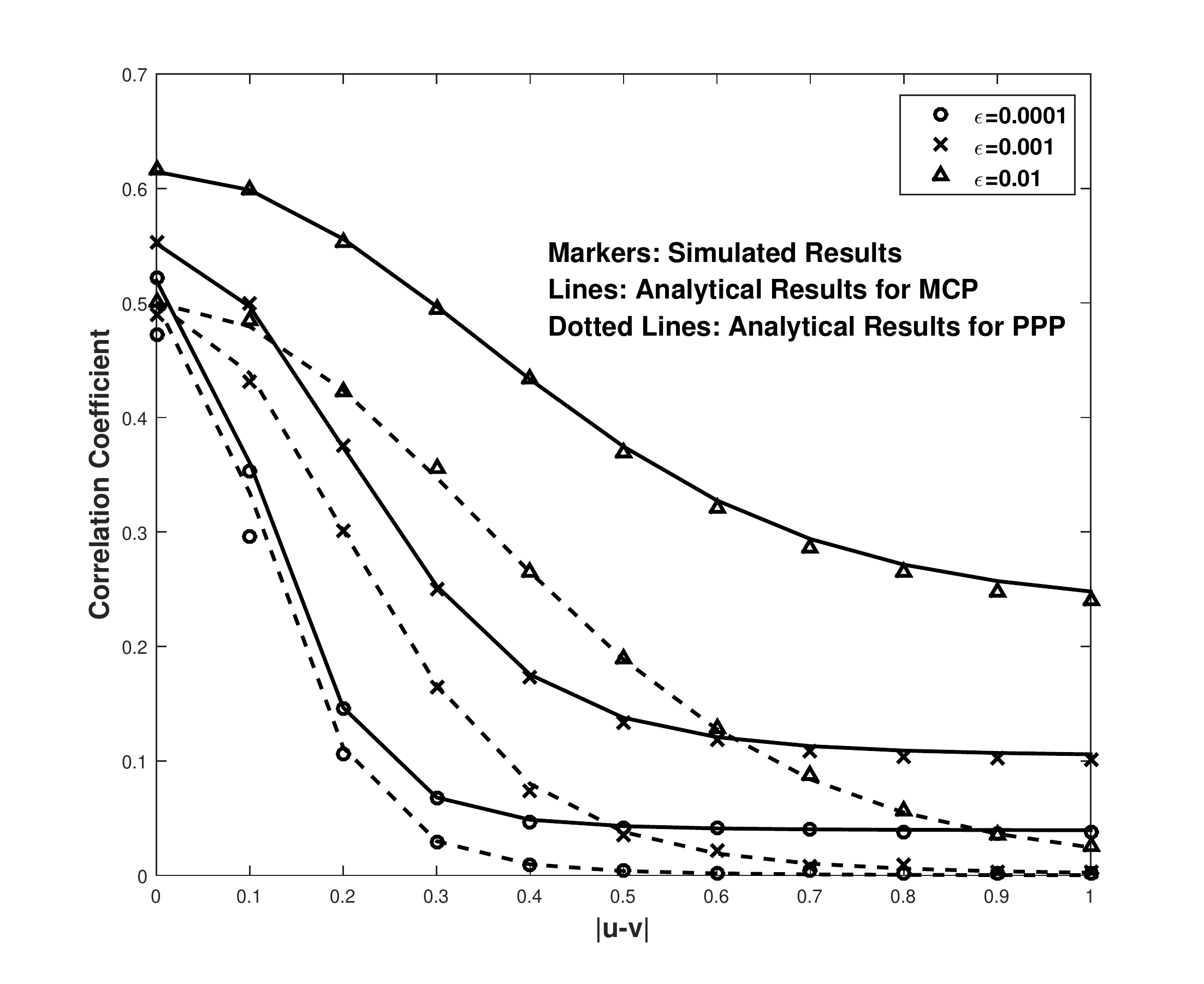}\caption{Interference correlation coefficient with different $\epsilon$ ($\alpha=4$,
$P=43$dBm, $\lambda_{p}=0.01,$ $c=3$, $R=1$)}

\centering
\end{figure}

\begin{figure}[tbh]
\centering\includegraphics[scale=0.4]{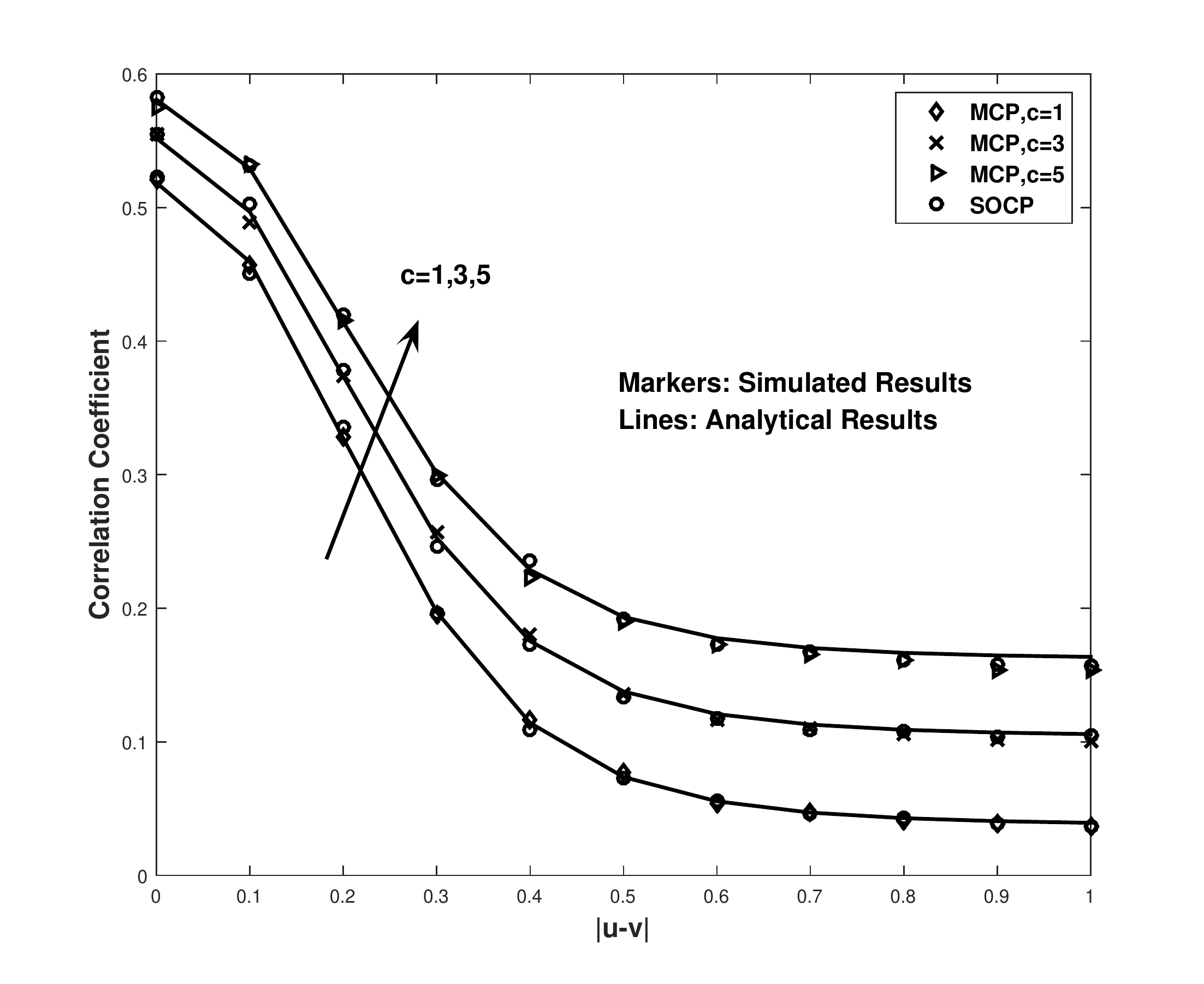}\caption{Interference correlation coefficient with different $c$ ($\alpha=4$,
$P=43$dBm, $\lambda_{0}=\lambda_{p}=0.01,$ $c_{1}=10$, $D_{1}=10$,
$c_{2}=c$, $D_{2}=R=1$, $\sigma=3$, $\epsilon=0.00001$)}
\end{figure}

\section{Performance Analysis}

In this section, Monte Carlo simulations are conducted to validate the
analysis of interference correlation coefficients. Further, we show
the impact of the average number of points and the radius of
each cluster on the interference correlation. In the simulation, the
interferers follow PPP, MCP, and SOCP. For comparing the interference
correlation coefficients under different distributed interferers, we
set the densities of interferers as $\lambda_{PPP}=\lambda_{p}c=\lambda_{0}c_{1}c_{2}$.

Fig. 3 compares the interference correlation coefficient in the case
of the interferers following MCP and PPP. The lines and dotted lines
represent the analytical results for MCP and PPP, respectively, while
the markers are the corresponding simulated results. First, our results in Theorem 1 are validated as the simulated results match well with
the analytical results. In addition, we find that the correlation
of interference in MCP is greater than that in PPP. It shows that
the spatial correlation in interferers has a significant effect on
interference correlation. Specifically, the attraction among the interferers
enhances the interference correlation. Moreover, as the distance
between the tested points grows, the interference correlation coefficient
decreases, which is coincide with our intuition. The farther the two
points are, the less interference correlation they have.

Fig. 4 illustrates the interference correlation coefficients under
different mean numbers of points in each cluster. Note that we only
show the analytical interference correlation coefficient for the interferers
following MCP. This is because, according to our analysis (Remark
1), the interference correlation coefficient under SOCP is the same
as that under MCP when $c=c_{2}$ and $D_{2}=R$. From Fig. 4, we
find that the interference correlation coefficient is improved with
the increase in the mean number of points in each cluster. The reason
is that the clustering in interferers exacerbates the interference
correlation. Given the radius of each cluster, increasing the average
number of each cluster aggravates the clustering in interferers. Thus,
increasing the mean number of points in each cluster enhances the
correlation in interference.

\begin{figure}[t]
\centering\includegraphics[scale=0.4]{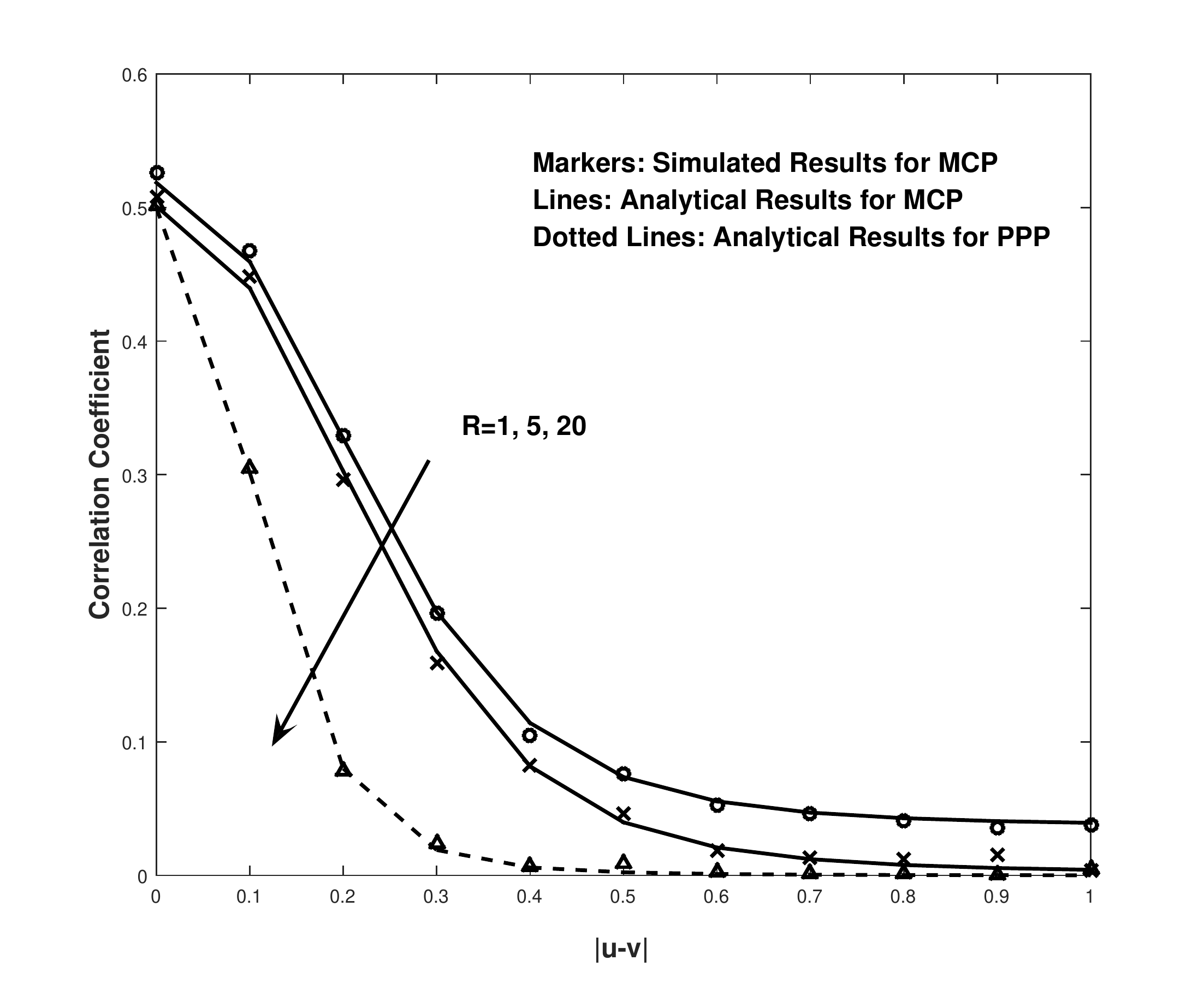}\caption{Interference correlation coefficient with different $R$ ($\alpha=4$,
$P=43$dBm, $\lambda_{0}=\lambda_{p}=0.01,$ $c_{1}=10$, $D_{1}=10$,
$c_{2}=c=1$, $D_{2}=R$, $\sigma=3$, $\epsilon=0.001$)}
\end{figure}

Fig. 5 shows the interference correlation coefficients varying with
different radiuses of each cluster. It is shown that the interference
correlation coefficient decreases with the increase in the radius
of each cluster $R$. The reason is that, given the average number of
each cluster, the larger the radius is, the less clustering impact
is. As a result, the interference correlation coefficient decreases.
In particular, the interference correlation coefficients under MCP
and SOCP approximate to that under PPP when $R=20$. This is because
when $R=20$ and $c=1$, $\frac{c}{\pi R^{2}}\rightarrow0$. Based
on the conclusion in Proposition 2, the impact of clustering on the
interference correlation can be ignored in this case.

\section{Conclusions}

In this paper, we have quantified the interference correlation in
non-Poisson networks including Matern cluster networks and second-order
cluster networks. It is shown that both the interference correlation
coefficients are greater than that in Poisson networks, i.e., $\zeta_{MCP}>\zeta_{PPP}$
and $\zeta_{SOCP}>\zeta_{PPP}$. This indicates the clustering in
MCP and SOCP has a positive contribution on the interference correlation.
Moreover, we have proved that the decrease of the mean number of points
in each cluster $c$ or the increase in the radius of each cluster
$R$ mitigates the interference correlation. In particular, we have
pointed that the effect of clustering on interference correlation
can be ignored under the condition that $\frac{c}{\pi^{2}R^{2}}\rightarrow0$.

The used methodology and achieved results clear the way to analyze
the interference correlation in non-Poisson networks. A follow-up
topic is to investigate the interference correlation in other non-Poisson
networks. In addition, the analysis in this paper can be extended
to the performance analysis of HCNs with dependence.

\section{Appendix: Proofs}

\subsection{Proof of Theorem 1}
\begin{IEEEproof}
First, we derive the correlation coefficient when the interference
comes from the BSs following MCP. In this case, the interference at
a randomly chosen user $u$ at time slot $t$ is given as
\begin{flalign}
I_{t}\left(u\right) & =\sum_{x\in\phi_{MCP}}P_{t}h_{xu}\left(t\right)g\left(x-u\right)\nonumber \\
 & =\sum_{y\in\phi_{p}}\sum_{x\in\phi^{\left[y\right]}}P_{t}h_{xu}\left(t\right)g\left(x-u\right),
\end{flalign}
where $\phi_{p}$ represents the parent process of MCP and $\phi^{\left[y\right]}$
denotes the cluster associated with parent point $y\in\phi_{p}$.

The mean interference is given by
\begin{flalign}
 & \mathbb{E}\left[I_{t}\left(u\right)\right]\nonumber \\
 & =\mathbb{E}\left[\sum_{y\in\phi_{p}}\sum_{x\in\phi^{\left[y\right]}}P_{t}h_{xu}\left(t\right)g\left(x-u\right)\right]\nonumber \\
 & \overset{\left(a\right)}{=}P_{t}\mathbb{E}\left[h\right]\lambda_{p}\int_{\mathbb{R}^{2}}\mathbb{E}\left[\sum_{x\in\phi^{\left[y\right]}}g\left(x-u\right)\right]\mathrm{d}y\nonumber \\
 & \overset{\left(b\right)}{=}P_{t}\mathbb{E}\left[h\right]\lambda_{p}c\int_{\mathbb{R}^{2}}\int_{\mathbb{R}^{2}}g\left(x-u-y\right)f_{cl}\left(x\right)\mathrm{d}x\mathrm{d}y\nonumber \\
 & =P_{t}\mathbb{E}\left[h\right]\lambda_{p}c\int_{\mathbb{R}^{2}}g\left(x-u\right)\int_{\mathbb{R}^{2}}f_{cl}\left(x+y\right)\mathrm{d}y\mathrm{d}x\nonumber \\
 & \overset{\left(c\right)}{=}P_{t}\mathbb{E}\left[h\right]\lambda_{p}c\int_{\mathbb{R}^{2}}g\left(x\right)\mathrm{d}x,\label{eq:MeanInterf_MCP}
\end{flalign}
where $\left(a\right)$ and $\left(b\right)$ come from Campbell-Mecke
Theorem, $\left(c\right)$ follows the fact that $\int_{\mathbb{R}^{2}}f_{cl}\left(x\right)\mathrm{d}x=1$.

The mean product of $I_{t_{1}}\left(u\right)$ and $I_{t_{2}}\left(v\right)$
is given by
\begin{flalign}
 & \mathbb{E}\left[I_{t_{1}}\left(u\right),I_{t_{2}}\left(v\right)\right]\nonumber \\
 & =\mathbb{E}\left[\sum_{x\in\phi_{MCP}}\!\!\!\!\!P_{t}h_{xu}\left(t_{1}\right)g\left(x-u\right)\sum_{y\in\phi_{MCP}}\!\!\!\!\!P_{t}h_{yv}\left(t_{2}\right)g\left(y-v\right)\right]\nonumber \\
 & =\mathbb{E}\left[\sum_{x\in\phi_{MCP}}P_{t}^{2}h_{xu}\left(t_{1}\right)h_{xv}\left(t_{2}\right)g\left(x-u\right)g\left(x-v\right)\right]\nonumber \\
 & +\mathbb{E}\left[\sum_{x,y\in\phi_{MCP}}^{x\neq y}P_{t}^{2}h_{xu}\left(t_{1}\right)h_{yv}\left(t_{2}\right)g\left(x-u\right)g\left(y-v\right)\right].\label{eq:Mean product_original}
\end{flalign}
 Let
\[
F=\mathbb{E}\left[\sum_{x\in\phi_{MCP}}P_{t}^{2}h_{xu}\left(t_{1}\right)h_{xv}\left(t_{2}\right)g\left(x-u\right)g\left(x-v\right)\right],
\]
 and
\[
Q=\mathbb{E}\left[\sum_{x,y\in\phi_{MCP}}^{x\neq y}P_{t}^{2}h_{xu}\left(t_{1}\right)h_{yv}\left(t_{2}\right)g\left(x-u\right)g\left(y-v\right)\right].
\]
Next, we calculate $F$ and $Q$, respectively.
\begin{flalign}
F & =\mathbb{E}\left[\sum_{x\in\phi_{MCP}}P_{t}^{2}h_{xu}\left(t_{1}\right)h_{xv}\left(t_{2}\right)g\left(x-u\right)g\left(x-v\right)\right]\nonumber \\
 & =P_{t}^{2}\mathbb{E}\left[h\right]^{2}\mathbb{E}\left[\sum_{y\in\phi_{p}}\sum_{x\in\phi^{\left[y\right]}}g\left(x-u\right)g\left(x-v\right)\right]\nonumber \\
 & \overset{\left(a\right)}{=}P_{t}^{2}\mathbb{E}\left[h\right]^{2}\lambda_{p}c\int_{\mathbb{R}^{2}}g\left(x-u\right)g\left(x-v\right)\mathrm{d}x,\label{eq:F}
\end{flalign}
where $\left(a\right)$ comes from Campbell-Mecke Theorem and the
fact that $\int_{\mathbb{R}^{2}}f_{cl}\left(x\right)\mathrm{d}x=1$.
\begin{flalign}
Q & =\mathbb{E}\left[\sum_{x,y\in\phi_{MCP}}^{x\neq y}P_{t}^{2}h_{xu}\left(t_{1}\right)h_{yv}\left(t_{2}\right)g\left(x-u\right)g\left(y-v\right)\right]\nonumber \\
 & =P_{t}^{2}\mathbb{E}\left[h\right]^{2}\mathbb{E}\left[\sum_{x,y\in\phi_{MCP}}^{x\neq y}g\left(x-u\right)g\left(y-v\right)\right]\nonumber \\
 & \overset{\left(a\right)}{=}P_{t}^{2}\mathbb{E}\left[h\right]^{2}\int_{\mathbb{R}^{2}}\int_{\mathbb{R}^{2}}g\left(x\right)g\left(y\right)\rho_{MCP}^{\left(2\right)}\left(x,y\right)\mathrm{d}x\mathrm{d}y\nonumber \\
 & \overset{\left(b\right)}{=}P_{t}^{2}\mathbb{E}\left[h\right]^{2}\left(\lambda_{p}c\right)^{2}\int_{\mathbb{R}^{2}}\int_{\mathbb{R}^{2}}g\left(x\right)g\left(y\right)\mathrm{d}x\mathrm{d}y\nonumber \\
 & +P_{t}^{2}\mathbb{E}\left[h\right]^{2}\lambda_{p}c\cdot F\left(c,R\right),\label{eq:Q}
\end{flalign}
where $\left(a\right)$ follows from $x=x-u$, $y=y-v$, $\left(b\right)$
comes from the second moment density of MCP given by \cite[p. 128]{StoGeoBook-Martin},
$F\left(c,R\right)=\frac{c}{\pi^{2}R^{4}}\int_{\mathbb{R}^{2}}\int_{\mathbb{R}^{2}}g\left(x\right)g\left(y\right)A_{R}\left(\Vert x-y\Vert\right)\mathrm{d}x\mathrm{d}y$,\emph{
and} $A_{R}\left(r\right)=2R^{2}\mathcal{\arccos}\left(\frac{r}{2R}\right)-r\sqrt{R^{2}-\frac{r^{2}}{4}},\;0\leq r\leq2R$,
\emph{and} 0 \emph{for} $r>2R$. Substituting (\ref{eq:F}) and (\ref{eq:Q})
into (\ref{eq:Mean product_original}), we get the mean product of
$I_{t_{1}}\left(u\right)$ and $I_{t_{2}}\left(v\right)$
\begin{flalign}
\mathbb{E}\left[I_{t_{1}}\left(u\right),I_{t_{2}}\left(v\right)\right] & =P_{t}^{2}\mathbb{E}\left[h\right]^{2}\lambda_{p}c\int_{\mathbb{R}^{2}}g\left(x-u\right)g\left(x-v\right)\mathrm{d}x\nonumber \\
 & +P_{t}^{2}\mathbb{E}\left[h\right]^{2}\left(\lambda_{p}c\right)^{2}\int_{\mathbb{R}^{2}}\int_{\mathbb{R}^{2}}g\left(x\right)g\left(y\right)\mathrm{d}x\mathrm{d}y\nonumber \\
 & +P_{t}^{2}\mathbb{E}\left[h\right]^{2}\lambda_{p}c\cdot F\left(c,R\right),\label{eq:MeanProd_MCP}
\end{flalign}

Similarly, the second moment of interference is given as
\begin{flalign}
 & \mathbb{E}\left[I_{t}^{2}\left(u\right)\right]\nonumber \\
 & =P_{t}^{2}\mathbb{E}\left[h^{2}\right]\lambda_{p}c\int_{\mathbb{R}^{2}}g^{2}\left(x\right)\mathrm{d}x\nonumber \\
 & +P_{t}^{2}\mathbb{E}\left[h\right]^{2}\left(\lambda_{p}c\right)^{2}\int_{\mathbb{R}^{2}}\int_{\mathbb{R}^{2}}g\left(x-u\right)g\left(y-v\right)\mathrm{d}x\mathrm{d}y\nonumber \\
 & +P_{t}^{2}\mathbb{E}\left[h\right]^{2}\lambda_{p}c\cdot F\left(c,R\right).\label{eq:SecondMoment_MCP}
\end{flalign}

Interference correlation coefficient is defined as
\begin{equation}
\zeta\left(I_{t_{1}}\left(u\right),I_{t_{2}}\left(v\right)\right)=\frac{\mathbb{E}\left[I_{t_{1}}\left(u\right),I_{t_{2}}\left(v\right)\right]-\mathbb{E}\left[I_{t_{1}}\left(u\right)\right]\mathbb{E}\left[I_{t_{2}}\left(v\right)\right]}{\sqrt{var\left(I_{t_{1}}\left(u\right)\right)}\cdot\sqrt{var\left(I_{t_{2}}\left(v\right)\right)}}.\label{eq:CorrCoef_Definition}
\end{equation}

Substituting (\ref{eq:MeanInterf_MCP}), (\ref{eq:MeanProd_MCP}),
and (\ref{eq:SecondMoment_MCP}) into (\ref{eq:CorrCoef_Definition}),
we derive the correlation coefficient of interference when the interferers
following MCP.

Next, we will calculate the correlation coefficient of interference
when the interferers following SOCP. The key to the calculation is
to derive the second moment density of SOCP $\rho_{SOCP}^{\left(2\right)}\left(x,y\right)$.
According to \cite[p. 127]{StoGeoBook-Martin}, there are two contributions
to the second moment density including the one from pairs of points
in different clusters and the one from pairs of points in the same
cluster. Since difference clusters in SOCP are independent, the second
moment density of SOCP is expressed as
\begin{equation}
\rho_{SOCP}^{\left(2\right)}\left(u,v\right)=\lambda^{2}+\mathbb{E}\left[\sum_{x\in\phi_{P_{0}}}\sum_{y\in\phi_{P_{1}^{\left[x\right]}}}h\left(u,v\mid x,y\right)\right],
\end{equation}
where $\lambda=\lambda_{0}c_{1}c_{2}$ denotes the intensity of SOCP,
$\phi_{P_{0}}$ represents the parent process with intensity $\lambda_{0}$,
$\phi_{P_{1}^{\left[x\right]}}$ is the first-order cluster associated
with parent point $x\in\phi_{P_{0}}$, $h\left(u,v\mid x,y\right)$
denotes the conditional second moment density.
\begin{flalign}
 & \mathbb{E}\left[\sum_{x\in\phi_{P_{0}}}\sum_{y\in\phi_{P_{1}^{\left[x\right]}}}h\left(u,v\mid x,y\right)\right]\nonumber \\
\overset{\left(a\right)}{=} & \mathbb{E}\left[\sum_{x\in\phi_{P_{0}}}\sum_{y\in\phi_{P_{1}^{\left[x\right]}}}c_{2}f_{cl_{2}}\left(u-y-x\right)c_{2}f_{cl_{2}}\left(v-y-x\right)\right]\nonumber \\
\overset{\left(b\right)}{=} & \lambda_{0}c_{1}\!\!\left(c_{2}\right)^{2}\!\!\!\int_{\mathbb{R}^{2}}\!\!\!\int_{\mathbb{R}^{2}}\!\!f_{cl_{1}}\left(y\right)\!\!f_{cl_{2}}\left(u-y-x\right)\!\!f_{cl_{2}}\left(v-y-x\right)\mathrm{d}x\mathrm{d}y\nonumber \\
= & \lambda_{0}c_{1}\left(c_{2}\right)^{2}\left(f_{cl_{2}}\star f_{cl_{2}}\right)\left(u-v\right)\int_{\mathbb{R}^{2}}f_{cl_{1}}\left(y\right)\mathrm{d}y\nonumber \\
\overset{\left(c\right)}{=} & \lambda_{0}c_{1}\left(c_{2}\right)^{2}\left(f_{cl_{2}}\star f_{cl_{2}}\right)\left(u-v\right)\nonumber \\
\overset{\left(d\right)}{=} & \lambda_{0}c_{1}\left(c_{2}\right)^{2}\cdot\frac{A_{D_{2}}\left(\Vert x-y\Vert\right)}{\pi^{2}D_{2}^{4}},
\end{flalign}
where $\left(a\right)$ follows from the independence of the points
in the same cluster, $\left(b\right)$ comes from Campbell-Mecke Theorem,
$\left(c\right)$ follows from the fact that $\int_{\mathbb{R}^{2}}f_{cl_{1}}\left(y\right)\mathrm{d}y=1$,
$\left(d\right)$ comes from the calculation of $\left(f_{cl_{2}}\star f_{cl_{2}}\right)\left(u-v\right)$
which is given in \cite{StoGeoBook-Martin} and $A_{D_{2}}\left(r\right)=2D_{2}^{2}\arccos\left(\frac{r}{2D_{2}}\right)-r\sqrt{D_{2}^{2}-\frac{r^{2}}{4}},\;0\leq r\leq2D_{2}$.

Based on the derived second moment density of SOCP, we get the correlation
coefficient of interference when the interferers following SOCP%
\footnote{We omit the corresponding proof for space limitation, since it is
a straightforward extension of the correlation coefficient derivation
for MCP. %
}.
\end{IEEEproof}
\bibliographystyle{IEEEtran}

\end{document}